\documentclass[aps,preprint,groupedaddress]{revtex4-2}

\usepackage{amssymb}
\usepackage{amsmath}
\usepackage{mathrsfs}

\DeclareMathOperator{\sign}{\mathop{sign}}

\usepackage{txfonts}
\usepackage{graphicx}

\begin{document}

\title{Frustrations on decorated triangular lattice in Ising model}

\author{F.A. Kassan-Ogly}

\author{A.V. Zarubin}
\email{Alexander.Zarubin@imp.uran.ru}

\affiliation{M.N. Mikheev Institute of Metal Physics of Ural Branch of Russian Academy of Sciences, S. Kovalevskoy street 18, Ekaterinburg, 620108, Russia}

\begin{abstract}
We study the frustration properties of the Ising model on a decorated triangular lattice with an arbitrary number of decorating spins on all lattice bonds in the framework of an exact analytical approach based on the Kramers--Wannier transfer matrix method. Expressions for the entropy, heat capacity, and spontaneous magnetization of the lattice are obtained, including the residual (zero-temperature) entropy and residual (zero-temperature) spontaneous magnetization of the system. The existence of magnetic frustrations in such a model and their influence on the behavior of the thermodynamic functions of the system are shown. The new and most important result of our study is related to the description of the possible coexistence of frustrations and long-range magnetic order in partially ordered spin systems.
\end{abstract}

\maketitle

\section{Introduction}
\label{sec:intro}

Every year there is a growing interest in the unusual properties of
spin systems with magnetic frustrations~\cite{Kassan-Ogly:2010:,Balents:2010,Lacroix:2011,Diep:2020,Kudasov:2012:,Vasiliev:2018:,Markina:2021:}.
This is because frustrated materials do not order down to the lowest
measured temperature, despite the existence of a large exchange interactions.

The phenomenon of frustrations in the Ising model on an antiferromagnetic
triangular lattice was first discovered by Wannier in 1950~\cite{Wannier:1950},
although the very concept of magnetic frustrations was introduced
by Toulouse only in 1977~\cite{Toulouse:1977:1}.

Since then, the phenomenon of magnetic frustrations has been studied
not only in the Ising model (for example see Refs. \cite{Giacomini:1987,Valverde:2008,Jurcisinova:2018}),
but also in a number of other basic models of magnetism, such as the
Potts model \cite{Qin:2014,Farnell:2018}, Heisenberg \cite{Suttner:2014,Natori:2019},
$XY$ \cite{Pires:2012}, $XXZ$ \cite{Yao:2008}, Hubbard \cite{Nourse:2021,Batista:2003}
and a number of others, on different lattices with different topologies.

It should be noted that the topology of a number of lattices and the
ratio of interactions in the system generates competition between
interactions of atomic spins, which leads to magnetic frustrations
\cite{Kassan-Ogly:2022,Zarubin:2019:,Kassan-Ogly:2019}.

Frustration properties of spin systems found in theoretical papers
and numerical experiments, occurring on two-dimensional triangular
lattices \cite{Coldea:2003,Shimizu:2003} or kagome lattices \cite{Huang:2021,Norman:2016},
drew attention to real magnetic layered systems.

Also of great interest is decorated spin lattices, which, in connection
with the Ising model, were originally proposed for consideration by
Syozi in 1951~\cite{Syozi:1951}. The concept of decorating a lattice
consists in introducing an additional (decorating) spin for each connection
between the nodes of the original lattice (nodal spins). Syozi introduced
a new transformation (the so-called decoration-iteration transformation),
with the help of which one can deduce the thermodynamic properties
of a decorated lattice from the properties of the original undecorated
lattice. Later on the decoration-iteration transformation has been
generalized to multiple decorations with an arbitrary number of extra
spins on each bond of the original lattice~\cite{Miyazima:1968,Syozi:1972,Fisher:1959}.

In a number of studies, exact solutions were obtained and a theoretical
analysis of models of real decorated spin systems was carried out,
in which their unusual properties of thermodynamic and magnetic functions
due to frustrations were found \cite{Kaneyoshi:1997,Jascur:1998,Strecka:2011,Strecka:2005,Torrico:2018,Oitmaa:2005,Galisova:2011,Si:2020}
(also see discussion in \cite{Kassan-Ogly:2022}).

Further in the paper, it is shown that in order to study the thermodynamic
and magnetic characteristics of decorated frustrated spin systems,
we obtained the principal eigenvalues in the canonical Kramers--Wannier
transfer matrix form in the Ising model for a decorated triangular
lattice, as well as the entropy, heat capacity, and spontaneous magnetization
of the system. The temperature dependences of the entropy, heat capacity
of the system, and spontaneous magnetization, as well as the behavioral
functions of zero-temperature (residual) entropy and zero-temperature
(residual) spontaneous magnetization, on the degree (multiplicity)
of decoration of the spin system were subjected to a detailed study.

\section{Thermodynamic functions of the model}

We will consider the Ising model \cite{Ising:1925} of a two-dimensional
spin system on a decorated triangular lattice the Hamiltonian of which
has the form
\begin{equation}
\mathscr{H}=-\sum_{i=1}^{3}\left(\sum_{n_{i}=1}^{N_{i}}J_{i}\sigma_{n_{i}}\sigma_{n_{i}+1}+\sum_{m_{i}=0}^{d_{i}}J_{i}^{\prime}\sigma_{m_{i}}\sigma_{m_{i}+1}\right),\label{eq:T1D:H}
\end{equation}
where $J_{i}$ is the parameter of the exchange interaction between
atomic spins at the nodes of the nearest neighbors of the original
lattice in the $i$~direction; $J_{i}^{\prime}$ is the exchange
interaction parameter both between neighboring decorating spins and
between neighboring decorating and nodal lattice spins in the $i$~direction;
the symbol $\sigma_{n}$ denotes the $z$-projection of the spin operator
of an atom of magnitude $\sigma=\pm1(\uparrow,\downarrow)$ located
at the node~$n$; and $N_{i}$ is the number of nodes of the spin
lattice in the $i$~direction; $d_{i}$ is the multiplicity of lattice
decoration in the $i$~direction.

\begin{figure}[tbh]
\centering \includegraphics[scale=1.33]{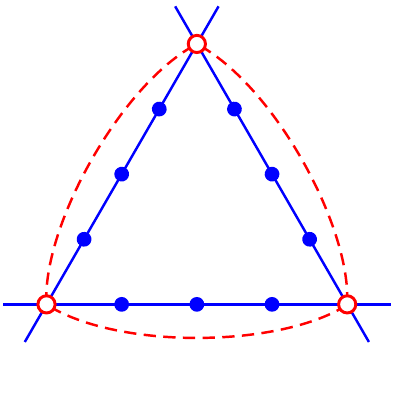}
\caption{Triangular lattice triply decorated in all directions. The empty red
circles mark the nodal spins, and the filled blue circles mark the
decorating spins. The dashed red lines indicate the interactions between
spins at the nodes of lattice ($J$), and the solid blue lines indicate
the interactions between decorating spins and also between decorating
and nodal spins ($J^{\prime}$).}
\label{fig:T1D:lattice}
\end{figure}

An example of such a decorated spin system is shown in Fig.~\ref{fig:T1D:lattice},
where the triangular triply decorated lattice in all directions are
demonstrated.

Using the decoration-iteration transformation approach proposed in
Ref.~\cite{Syozi:1951}, we obtained analytical expressions for the
principal (the only one maximal positive real) eigenvalue of the Kramers--Wannier
transfer matrix ($\lambda$) in the Ising model on a decorated triangular
lattice
\begin{multline}
\ln\frac{\lambda}{2}=\frac{1}{8\pi^{2}(d_{1}+d_{2}+d_{3}+1)}\intop_{0}^{2\pi}\intop_{0}^{2\pi}\ln[C_{1}C_{2}C_{3}+S_{1}S_{2}S_{3}\\
-S_{1}D_{2}D_{3}\cos\alpha-D_{1}S_{2}D_{3}\cos\beta-D_{1}D_{2}S_{3}\cos(\alpha+\beta)]\:d\alpha\:d\beta,\label{eq:T1D:L1}
\end{multline}
where
\begin{equation}
C_{i}=\frac{1}{2}\left[e^{\frac{2J_{i}}{T}}\left(c_{id}^{1+d_{i}}+s_{id}^{1+d_{i}}\right)^{2}+e^{-\frac{2J_{i}}{T}}\left(c_{id}^{1+d_{i}}-s_{id}^{1+d_{i}}\right)^{2}\right],\label{eq:T1D:L1:C}
\end{equation}
\begin{equation}
S_{i}=\frac{1}{2}\left[e^{\frac{2J_{i}}{T}}\left(c_{id}^{1+d_{i}}+s_{id}^{1+d_{i}}\right)^{2}-e^{-\frac{2J_{i}}{T}}\left(c_{id}^{1+d_{i}}-s_{id}^{1+d_{i}}\right)^{2}\right],\label{eq:T1D:L1:S}
\end{equation}
\begin{equation}
D_{i}=\left(c_{id}^{1+d_{i}}\right)^{2}-\left(s_{id}^{1+d_{i}}\right)^{2},\label{eq:T1D:L1:D}
\end{equation}
\[
c_{id}=\cosh\frac{J_{i}^{\prime}}{T},\quad s_{id}=\sinh\frac{J_{i}^{\prime}}{T},\quad i=1,\ 2,\ 3,
\]
here $T$ is the absolute temperature.

Hereinafter, the Boltzmann constant ($k_{\text{B}}$) will be set
equal to unity, and the quantities: $T$, $J^{\prime}$ will be measured
in $|J|$ units, as is done in the theory of low-dimensional systems.

It is easy to see that the expression (\ref{eq:T1D:L1}) for the principal
eigenvalue of the transfer matrix is determined by the geometry of
the lattice and the number of nodal and decorating exchange interactions.

Thus, knowing the principal eigenvalue of the transfer-matrix (\ref{eq:T1D:L1}),
we can calculate the Helmholtz free energy of the system per spin
\begin{equation}
F=-T\ln\lambda.\label{eq:F}
\end{equation}
From here one can find all the thermodynamic functions of the system,
such as the specific entropy of the system
\begin{equation}
s=-\frac{\partial F}{\partial T}=\ln\lambda+\frac{T}{\lambda}\frac{\partial\lambda}{\partial T}\label{eq:s}
\end{equation}
and the magnetic specific heat capacity
\begin{equation}
c=-T\frac{\partial^{2}F}{\partial T^{2}}=2\frac{T}{\lambda}\frac{\partial\lambda}{\partial T}+\frac{T^{2}}{\lambda}\frac{\partial^{2}\lambda}{\partial T^{2}}-\frac{T^{2}}{\lambda^{2}}\left(\frac{\partial\lambda}{\partial T}\right)^{2}.\label{eq:c}
\end{equation}

An expression for the spontaneous magnetization was also obtained,
which is defined as the square root of pairwise spin-spin correlation
function at distance between spins tending to infinity,
\begin{equation}
M=\sqrt{\lim_{\Delta\to\infty}\langle\sigma_{1}\sigma_{1+\Delta}\rangle},\label{eq:M}
\end{equation}
(see, for example Refs.~\cite{Montroll:1963,Baxter:2011}).

We obtained the expression for the spontaneous magnetization on a
decorated triangular lattice in most convenient form
\begin{equation}
M=(1-k)^{1/8},\quad k=\frac{D_{1}^{2}D_{2}^{2}D_{3}^{2}}{S_{1}^{2}S_{2}^{2}D_{3}^{2}+S_{1}^{2}D_{2}^{2}S_{3}^{2}+D_{1}^{2}S_{2}^{2}S_{3}^{2}+2S_{1}^{2}S_{2}^{2}S_{3}^{2}+2C_{1}C_{2}C_{3}S_{1}S_{2}S_{3}}.\label{eq:T1D:M}
\end{equation}

\section{Thermodynamics of the spin system without decorations}

Even without taking into account the lattice decorations ($J^{\prime}=0$),
the model under consideration contains eight options for the ratios
of the parameters of exchange interactions of spins at the nodes of
the nearest neighbors of the triangular lattice
\[
(J_{1}>0,J_{2}>0,J_{3}>0),\quad(J_{1}>0,J_{2}<0,J_{3}<0),
\]
\begin{equation}
(J_{1}<0,J_{2}>0,J_{3}<0),\quad(J_{1}<0,J_{2}<0,J_{3}>0),\label{eq:T1D:C85:C32}
\end{equation}
\[
(J_{1}<0,J_{2}<0,J_{3}<0),\quad(J_{1}<0,J_{2}>0,J_{3}>0),
\]
\begin{equation}
(J_{1}>0,J_{2}<0,J_{3}>0),\quad(J_{1}>0,J_{2}>0,J_{3}<0).\label{eq:T1D:C14:C67}
\end{equation}
The first four sets (\ref{eq:T1D:C85:C32}) correspond to such types
of interactions in which there is no competition between exchange
interactions in the spin system. The last four variants of the parameters
(\ref{eq:T1D:C14:C67}) define a system with competing exchange interactions
between spins.

In the absence of decorations ($J^{\prime}=0$) in the spin system,
the expressions (\ref{eq:T1D:L1:C})--(\ref{eq:T1D:L1:D})
determined by the principal eigenvalue of the transfer matrix (\ref{eq:T1D:L1}),
take the form
\[
C_{i}=\cosh\frac{2J_{i}}{T},\quad S_{i}=\sinh\frac{2J_{i}}{T},\quad D_{i}=1,\quad i=1,\ 2,\ 3,
\]
and spontaneous magnetization (\ref{eq:T1D:M}) has the form
\begin{equation}
M=(1-k)^{1/8},\quad k=\frac{1}{S_{1}^{2}S_{2}^{2}+S_{1}^{2}S_{3}^{2}+S_{2}^{2}S_{3}^{2}+2S_{1}^{2}S_{2}^{2}S_{3}^{2}+2C_{1}C_{2}C_{3}S_{1}S_{2}S_{3}},\label{eq:T1D:M:1}
\end{equation}
which is the same as the solution in Ref.~\cite{Syozi:1960}.

Also note that the partition function of the Ising model on any two-dimensional
lattice is the sum of $2^{N}$ configurations with all possible fixed
spin values ($+1$ or $-1$) with corresponding Boltzmann statistical
weights ($W$). At an infinite temperature, the statistical weights
of all configurations are equal, the system is completely degenerate,
and the entropy of the system is equal to the natural logarithm of
the statistical weight of the state,
\begin{equation}
\lim_{T\to\infty}s\equiv s^{\infty}=\ln2\approx0.693\,147.\label{eq:s:T:infy}
\end{equation}
In this case, the statistical weight is equal to the number of states
on the node ($W=2$) in the considered Ising model.

As the temperature decreases and depending on the ratios of the model
parameters, the degeneracy of the spin system can decrease and decrease
down to zero temperature. At the same time, the statistical weights
of the system configurations decrease unequally, continuously reducing
the entropy of the system.

Depending on the specific set of parameters of exchange interactions
between atomic spins, several situations of existence of various magnetic
states can arise in the system, which are determined by the features
of the ordering of the spin system.

1. In the absence of competing interactions between the nodal spins
of the triangular lattice, that is, with the sets of exchange interactions
(\ref{eq:T1D:C85:C32}), in the ground state ($T=0$) only one spin
configuration ($W=1$) is retained, the system degeneracy disappears
and a magnetically ordered equilibrium state appears in the system.
The zero-temperature (residual) entropy of such a spin system is zero,
\[
s(T=0)\equiv s^{\circ}=\ln1=0,
\]
which corresponds to the Nernst--Planck theorem~\cite{Sommerfeld:1956}.

In this situation, the retained configuration has translational invariance,
which means the existence of a long-range magnetic order in the ground
state of the system, while the zero-temperature (residual) spontaneous
magnetization reaches saturation, i.e.
\[
M(T=0)\equiv M^{\circ}=1.
\]

\begin{figure}[tbh]
\centering \includegraphics[scale=1.33]{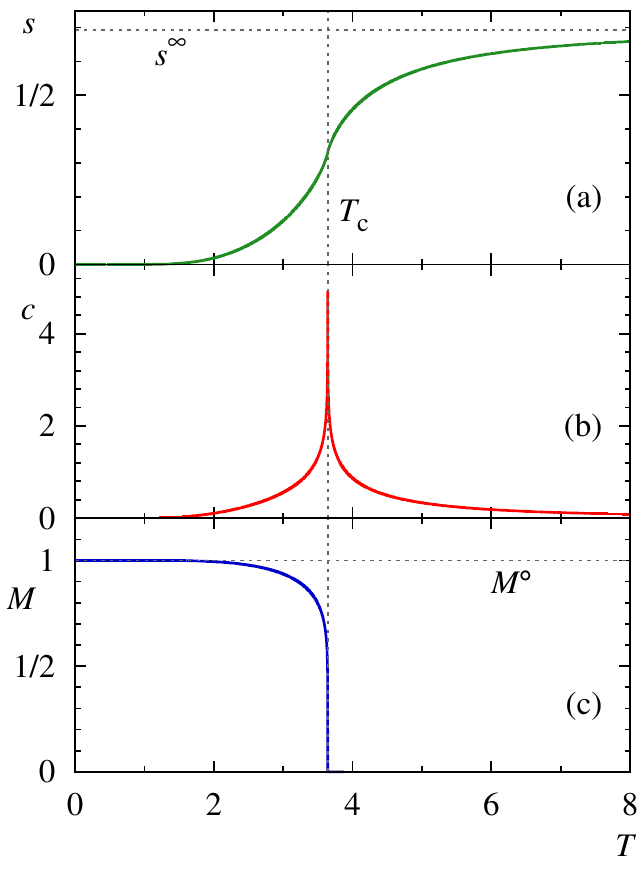}
\caption{Entropy (green line, a), heat capacity (red line, b) and spontaneous
magnetization (blue line, c) for an undecorated triangular lattice
at antiferro-antiferro-ferromagnetic nodal spin exchange interactions
($J_{1}=J_{2}=-J_{3}=-1$), where $s^{\circ}=0$, $T_{\text{c}}\approx3.640\,957$,
$M^{\circ}=1$.}
\label{fig:T1D:mmp:000:TD}
\end{figure}

The temperature evolutions of the heat capacity and spontaneous magnetization
of the system demonstrate a magnetic phase transition at a nonzero
temperature ($T_{\text{c}}>0$). The heat capacity of the system experiences
a logarithmic divergence ($\lambda$-shaped Onsager peak) at this
critical point ($T_{\text{c}}$). The spontaneous magnetization (\ref{eq:T1D:M:1})
decreases with increasing temperature and vanishes at temperatures
above the critical point ($T>T_{\text{c}}$). This behavior of these
thermodynamic functions confirms the existence of long-range magnetic
order at low temperatures ($T<T_{\text{c}}$). This situation is shown
in Fig.~\ref{fig:T1D:mmp:000:TD}.

It is important to note that hereinafter we will only consider the
cases where 
\begin{equation}
|J_{i}|=1.\label{eq:J:1}
\end{equation}
With such parameters of the exchange interaction between atomic spins,
one should expect simplicity and clarity in the consideration of the
thermodynamic functions of the spin system, as well as the appearance
of competition between the exchange interactions of lattice spins.

In this case, the phase-transition temperature is
\[
T_{\text{c}}=\frac{4}{\ln3}\approx3.640\,957.
\]

2. On the other hand, in the cases of the existence of competing exchange
interactions between the nodal spins of the triangular lattice, with
the ratio of the model parameters (\ref{eq:T1D:C14:C67}), in the
ground state the spin system does not experience magnetic ordering,
the system is degenerate even in the ground state, while the statistical
weight is greater than unity ($W>1$), and the residual entropy of
the system is greater than zero,
\begin{equation}
s^{\circ}=\ln W>0.\label{eq:s0:fr}
\end{equation}
Therefore, such a degenerate state of the system, in which the statistical
weight of the ground state of the system is in the range $1<W\leqslant2$,
and the entropy is greater than zero, $0<s^{\circ}\leqslant s^{\infty}$,
should be called \emph{frustrated}. Note that the existence of a non-zero
residual entropy does not contradict the third law of thermodynamics~\cite{Sommerfeld:1956}
and was discussed in detail in Refs.~\cite{Zarubin:2019:,Kassan-Ogly:2022}.

In other words, even when the temperature is lowered, the competition
between exchange interactions does not allow for magnetic ordering
in the spin system, which leads to complete magnetic disorder even
in the ground state. At the same time, a set of spin configurations
comparable with the size of the system with the same statistical weights,
including those with the absence of translational invariance, is retained,
the degeneracy does not disappear even at zero temperature, and therefore
the residual entropy of such a system is nonzero (\ref{eq:s0:fr}).

As the simplest and most understandable explanation of this unusual
situation, one can give examples of one-dimensional Ising systems
discussed in Refs.~\cite{Zarubin:2019:,Zarubin:2020}.

The presence of a nonzero residual entropy indicates that the spin
system is in the regime of magnetic frustrations. Frustrations destroy
the long-range magnetic order, as a result of which there is no magnetic
phase transition in the system. The temperature dependence of the
heat capacity has a single dome-shape peak that does not diverge,
which indicates the absence of a magnetic phase transition at finite
temperatures. In turn, such a system does not exhibit spontaneous
magnetization at any temperature.

The fully antiferromagnetic triangular lattice Ising model was discussed
in the original papers~\cite{Wannier:1950,Houtappel:1950}.

This type of behavior of undecorated spin systems is observed, for
example, in cases of antiferro-antiferro-antiferromagnetic interactions
between nodal spins ($J_{1}=J_{2}=J_{3}=-1$), as well as in one antiferromagnetic
and two ferromagnetic interactions ($J_{i}=-1,J_{j}=J_{k}=+1$) when
the residual entropy of the system is finite and equal to
\begin{equation}
s^{\circ}=\frac{2}{\pi}\intop_{0}^{\pi/3}\ln(2\cos\alpha)\,d\alpha\approx0.323\,066.\label{eq:T1D:S0:Wan}
\end{equation}
Note here that in Wannier's original paper~\cite{Wannier:1950} an
error was received when calculating (\ref{eq:T1D:S0:Wan}).

\begin{figure}[tbh]
\centering \includegraphics[scale=1.33]{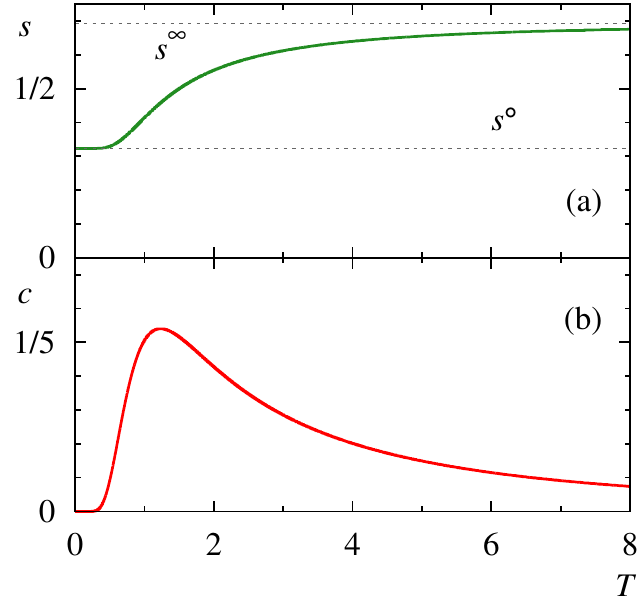}
\caption{Entropy (green line, a) and heat capacity (red line, b) for an undecorated
triangular lattice in one of the variants of competing exchange interactions,
at ferro-ferro-antiferromagnetic exchange interactions of nodal spins
($J_{i}=J_{j}=-J_{k}=+1$), where $s^{\circ}\approx0.323\,066$, $T_{\text{c}}=0$,
$M^{\circ}=0$.}
\label{fig:T1D:ppm:000:TD}
\end{figure}

This situation is shown in Fig.~\ref{fig:T1D:ppm:000:TD}.

3. We should also mention cases where one or more exchange interactions
between nodal spins are equal to zero.

First, with one zero exchange interaction between nodal spins in a
triangular lattice and in the presence of two other non-zero interactions
(in this case, the spin system is equivalent to a system on a square
lattice), 
\[
|J_{i}|>0,\quad|J_{j}|>0,\quad J_{k}=0,
\]
the spin system in the ground state is not degenerate and experiences
magnetic ordering, therefore the residual entropy of the system is
zero, and the residual spontaneous magnetization is equal to unity,
\[
s^{\circ}=0,\quad M^{\circ}=1.
\]
In this case, the heat capacity of the system has a $\lambda$-shaped
Onsager peak at the phase-transition temperature, which is equal to
\[
T_{\text{c}}=\frac{2}{\ln(1+\sqrt{2})}\approx2.269\,185
\]
(which is the same as the phase-transition temperature for a square
lattice obtained by Onsager in 1944 \cite{Onsager:1944}), and the
spontaneous magnetization decreases with increasing temperature and
becomes zero above this phase-transition temperature.

In this case, the expression for spontaneous magnetization (\ref{eq:T1D:M:1})
is converted to the following form 
\begin{equation}
M_{\text{Ons}}=\left[1-\left(\sinh\frac{2J_{1}}{T}\right)^{-2}\left(\sinh\frac{2J_{2}}{T}\right)^{-2}\right]^{1/8}.\label{eq:T1D:M:Ons}
\end{equation}

It is important to note here that, in fact, the beginning of the study
of spontaneous magnetization in the Ising model on two-dimensional
lattices was also laid by Onsager on a square lattice in 1948 \cite{Baxter:2011},
when at the conference at Cornell University he simply wrote his famous
formula on a blackboard with chalk (\ref{eq:T1D:M:Ons}).

Following Onsager, the exact solution for the free energy in the Ising
model on a triangular lattice was obtained by Wannier in 1950 \cite{Wannier:1950},
after which the study of spontaneous magnetization in the Ising model
was undertaken by various methods and authors, in particular by Potts
\cite{Potts:1952}
\begin{equation}
M_{\text{P}}=(1-k_{\text{P}})^{1/8},\label{eq:T1D:M:Potts}
\end{equation}
\[
k_{\text{P}}=\frac{16x_{1}^{2}x_{2}^{2}x_{3}^{2}}{(1+x_{1}x_{2}+x_{2}x_{3}+x_{3}x_{1})(1+x_{1}x_{2}-x_{2}x_{3}-x_{3}x_{1})(1-x_{1}x_{2}+x_{2}x_{3}-x_{3}x_{1})(1-x_{1}x_{2}-x_{2}x_{3}+x_{3}x_{1})},
\]
\[
x_{i}=e^{-\frac{2J_{i}}{T}},\quad i=1,\ 2,\ 3,
\]
Stephenson \cite{Stephenson:1964:T1}
\begin{equation}
M_{\text{S}}=(1-k_{\text{S}})^{1/8},\label{eq:T1D:M:Stephenson}
\end{equation}
\[
k_{\text{S}}=\frac{\left[(1-t_{1}^{2})(1-t_{2}^{2})(1-t_{3}^{2})\right]^{2}}{16(1+t_{1}t_{2}t_{3})(t_{1}+t_{2}t_{3})(t_{2}+t_{3}t_{1})(t_{3}+t_{1}t_{2})},
\]
\[
t_{i}=\tanh\frac{J_{i}}{T},\quad i=1,\ 2,\ 3,
\]
Syozi and Naya \cite{Syozi:1960}
\begin{equation}
M_{\text{SN}}=(1-k_{\text{SN}})^{1/8},\label{eq:T1D:M:Syozi:Naya}
\end{equation}
\[
k_{\text{SN}}=\frac{1}{(s_{1}s_{2})^{2}+(s_{2}s_{3})^{2}+(s_{3}s_{1})^{2}+2(s_{1}s_{2}s_{2})^{2}+2c_{1}c_{2}c_{3}s_{1}s_{2}s_{3}},
\]
\[
s_{i}=\sinh\frac{2J_{i}}{T},\quad c_{i}=\cosh\frac{2J_{i}}{T},\quad i=1,\ 2,\ 3.
\]
All expressions for spontaneous magnetization obtained in these papers
are presented in the same way as in the formulas for a square lattice,
through hyperbolic sines and cosines, exponentials, and hyperbolic
tangents.

We must note here that our expression (\ref{eq:T1D:M:1}) and all
the above expressions (\ref{eq:T1D:M:Potts}), \ref{eq:T1D:M:Stephenson})
and (\ref{eq:T1D:M:Syozi:Naya}) are equivalent to each other.

\begin{figure}[tbh]
\centering \includegraphics[scale=1.33]{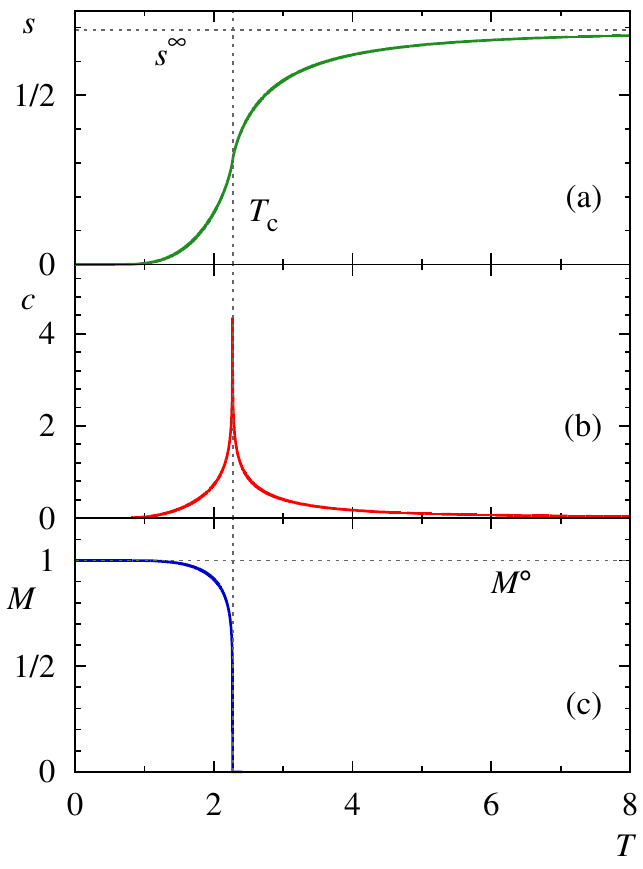}
\caption{Entropy (green line, a), heat capacity (red line, b) and spontaneous
magnetization (blue line, c) for an undecorated triangular lattice
with two nonzero and one zero nodal spin exchange interactions ($J_{i}=-J_{j}=+1$,
$J_{k}=0$), where $s^{\circ}=0$, $T_{\text{c}}\approx2.269\,185$,
$M^{\circ}=1$.}
\label{fig:T1D:pm0:000:TD}
\end{figure}

In the case under consideration, the behavior of the thermodynamic
functions is shown in Fig.~\ref{fig:T1D:pm0:000:TD}, which is equivalent
to the behavior of the functions shown in Fig. \ref{fig:T1D:mmp:000:TD}.

\begin{figure}[tbh]
\centering \includegraphics[scale=1.33]{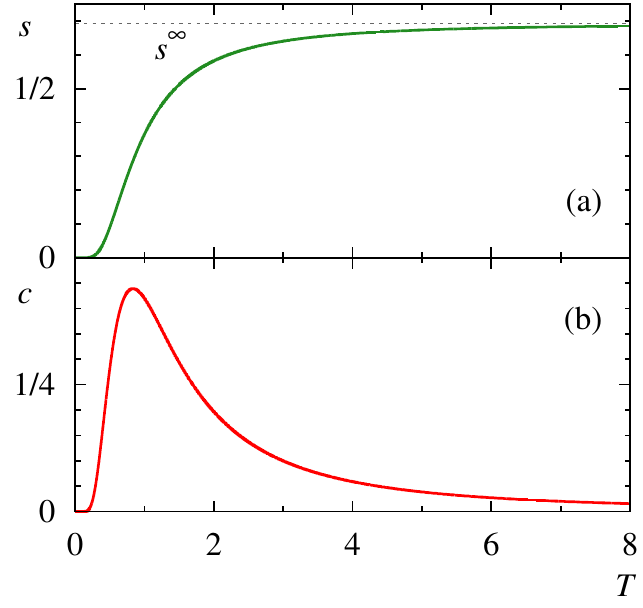}
\caption{Entropy (green line, a) and heat capacity (red line, b) for an undecorated
triangular lattice with a single non-zero (antiferromagnetic) and
other zero nodal spin exchange interactions ($J_{i}=-1$, $J_{j}=J_{k}=0$),
where $s^{\circ}=0$, $T_{\text{c}}=0$, $M^{\circ}=0$.}
\label{fig:T1D:m00:000:TD}
\end{figure}

4. Second, with two zero exchange interactions between nodal spins
in a triangular lattice (in this case, the spin system is equivalent
to a set of 1D spin lattices), 
\[
|J_{i}|>0,\quad J_{j}=J_{k}=0,
\]
the residual entropy of the system is zero. The magnetic phase-transition
temperature is also equal to zero ($T_{\text{c}}=0$), while the temperature
evolution of the heat capacity of the system has single-domed peak,
and there is no spontaneous magnetization, as shown in Fig. \ref{fig:T1D:m00:000:TD}.

5. At the end of this section, as a special case, we can consider
situations where all exchange interactions between nodal spins are
zero ($J_{i}=0$). In such a situation, a paramagnetic state of the
system is realized, characterized by the fact that all configurations
of the system have the same probability and have the same energy at
all temperatures. The entropy of such a state of the system is equal
to the natural logarithm of two,
\begin{equation}
s=\ln2\approx0.693\,147,\label{eq:T1D:S0:0}
\end{equation}
and it is the same (maximum) at any temperature. From this it is clear
that \emph{the Ising paramagnet is an absolutely frustrated system}
(see the discussion in Refs.~\cite{Zarubin:2019:,Zarubin:2020}).

\section{Residual entropy of a decorated spin system}

It is obvious that when the decoration of the spin system ($|J^{\prime}|>0$)
is taken into account, the situation becomes more complicated, including
the emergence of multiple competing exchange interactions between
nodal and decorating spins, which can generate magnetic frustrations
in such systems. The frustrations themselves can suppress the magnetic
ordering of the system, forming a degenerate state.

In continuation of the logic stated in the relation (\ref{eq:J:1}),
we will further consider fairly simple and illustrative examples for
the case when the values of exchange interactions between nodal and
between decorating spins are equal in absolute value to unity,
\begin{equation}
|J_{i}|=1,\quad|J_{i}^{\prime}|=1.\label{eq:JJ:1}
\end{equation}
It is this situation that will make it possible to visually demonstrate
the competition between exchange interactions between spins for their
equal values.

As noted earlier, the presence or absence of frustrations in a spin
system can be determined by calculating the residual entropy.

Note that from the formula for the entropy (\ref{eq:s}) one can obtain
expressions for the residual entropy of the system in the form
\begin{equation}
s^{\circ}=\frac{1}{8\pi^{2}(d_{1}+d_{2}+d_{3}+1)}\intop_{0}^{2\pi}\intop_{0}^{2\pi}\ln[R_{0}+R_{1}\cos\alpha+R_{2}\cos\beta+R_{3}\cos(\alpha+\beta)]\:d\alpha\:d\beta,\label{eq:T1D:s0}
\end{equation}
where the coefficients of the expression depend on the parameters
of the exchange interactions between nodal and decorating spins, as
well as on the multiplicity of decorating the system.

In the following paragraphs, we calculate the values of the residual
entropy of the system for various ratios of the model parameters.

1. In the presence of antiferromagnetic interactions between nodal
and decorating spins,
\begin{equation}
J_{i}=-1,\quad J_{j}^{\prime}=-1,\quad i,j=1,\ 2,\ 3\label{eq:T1D:Jm:Jm}
\end{equation}
there are three cases where the residual entropy (\ref{eq:T1D:s0})
is non-zero.

First, for all even values of decoration multiplicity,
\[
d_{i}=2k,\quad k\in\mathbb{Z},\quad k\geqslant0,\quad i=1,\ 2,\ 3
\]
the coefficients of the expression for the residual entropy (\ref{eq:T1D:s0})
are 
\begin{equation}
R_{0}=\frac{1+(-1)^{d_{1}}}{2}(d_{1}+1)^{2}+\frac{1+(-1)^{d_{2}}}{2}(d_{2}+1)^{2}+\frac{1+(-1)^{d_{3}}}{2}(d_{3}+1)^{2},\label{eq:T1D:R0:1.1}
\end{equation}
\begin{align}
R_{1}&=(1+(-1)^{d_{1}})(d_{2}+1)(d_{3}+1),\notag\\
R_{2}&=(1+(-1)^{d_{2}})(d_{1}+1)(d_{3}+1),\notag\\
R_{3}&=(1+(-1)^{d_{3}})(d_{1}+1)(d_{2}+1).\label{eq:T1D:Rx:1.1}
\end{align}

Secondly, for all odd values of decoration multiplicity,
\[
d_{i}=2k+1,\quad k\in\mathbb{Z},\quad k\geqslant0,\quad i=1,\ 2,\ 3
\]
the coefficients of the expression for the residual entropy (\ref{eq:T1D:s0})
are
\begin{equation}
R_{0}=\frac{1}{2}[(d_{1}^{2}+2d_{1}+2)(d_{2}^{2}+2d_{2}+2)(d_{3}^{2}+2d_{3}+2)-d_{1}(d_{1}+2)d_{2}(d_{2}+2)d_{3}(d_{3}+2)],\label{eq:T1D:R0:1.2}
\end{equation}
\begin{align}
R_{1}&=2d_{1}(d_{1}+2)(d_{2}+1)(d_{3}+1),\notag\\
R_{2}&=2(d_{1}+1)d_{2}(d_{2}+2)(d_{3}+1),\notag\\
R_{3}&=2(d_{1}+1)(d_{2}+1)d_{3}(d_{3}+2).\label{eq:T1D:Rx:1.2}
\end{align}

Third, for two odd 
\[
d_{i}=2k+1,\quad k\in\mathbb{Z},\quad k\geqslant0
\]
and one even 
\[
d_{j}=2k,\quad k\in\mathbb{Z},\quad k\geqslant0
\]
values of the decoration multiplicity, the coefficients of the expression
for the residual entropy (\ref{eq:T1D:s0}) are
\begin{equation}
R_{0}=\frac{1-(-1)^{d_{1}}}{2}(d_{1}+1)^{2}+\frac{1-(-1)^{d_{2}}}{2}(d_{2}+1)^{2}+\frac{1-(-1)^{d_{3}}}{2}(d_{3}+1)^{2},\label{eq:T1D:R0:1.3}
\end{equation}
and the following coefficients $R_{1}$, $R_{2}$, and $R_{3}$ are
respectively (\ref{eq:T1D:Rx:1.1}).

In other cases, that is, with two even and one odd values of the decoration
multiplicity, the residual entropy is zero, and the coefficients of
the expression (\ref{eq:T1D:s0}) are equal to
\begin{equation}
R_{0}=1,\quad R_{1}=R_{2}=R_{3}=0.\label{eq:T1D:R:0}
\end{equation}

2. In the presence of antiferromagnetic interactions between nodal
spins and ferromagnetic interactions between decorating spins,
\begin{equation}
J_{i}=-1,\quad J_{j}^{\prime}=+1,\quad i,j=1,\ 2,\ 3\label{eq:T1D:Jm:Jp}
\end{equation}
for any values of decoration multiplicity 
\[
d_{i}=k,\quad k\in\mathbb{Z},\quad k\geqslant0,\quad i=1,\ 2,\ 3
\]
the residual entropy of the system is not equal to zero and the coefficients
of the expression for the residual entropy of the system (\ref{eq:T1D:s0})
are respectively (\ref{eq:T1D:R0:1.2}) and (\ref{eq:T1D:Rx:1.2}).

3. In the presence of ferromagnetic interactions between nodal spins
and antiferromagnetic interactions between decorating spins,
\begin{equation}
J_{i}=+1,\quad J_{j}^{\prime}=-1,\quad i,j=1,\ 2,\ 3\label{eq:T1D:Jp:Jm}
\end{equation}
there are three cases when the residual entropy of the system is not
equal to zero.

First, for all even values of decoration multiplicity, 
\[
d_{i}=2k,\quad k\in\mathbb{Z},\quad k\geqslant0,\quad i=1,\ 2,\ 3
\]
the coefficients of the expression for the residual entropy of the
system (\ref{eq:T1D:s0}) are
\begin{equation}
R_{0}=\frac{1}{2}[(d_{1}^{2}+2d_{1}+2)(d_{2}^{2}+2d_{2}+2)(d_{3}^{2}+2d_{3}+2)+d_{1}(d_{1}+2)d_{2}(d_{2}+2)d_{3}(d_{3}+2)],\label{eq:T1D:R0:3.1}
\end{equation}
\begin{align}
R_{1}&=-2d_{1}(d_{1}+2)(d_{2}+1)(d_{3}+1),\notag\\
R_{2}&=-2(d_{1}+1)d_{2}(d_{2}+2)(d_{3}+1),\notag\\
R_{3}&=-2(d_{1}+1)(d_{2}+1)d_{3}(d_{3}+2).\label{eq:T1D:Rx:3.1}
\end{align}

Second, for two even 
\[
d_{i}=2k,\quad k\in\mathbb{Z},\quad k\geqslant0
\]
and one odd 
\[
d_{j}=2k+1,\quad k\in\mathbb{Z},\quad k\geqslant0
\]
values of the decoration multiplicity, the coefficients of the expression
for the residual entropy of the system (\ref{eq:T1D:s0}) are equal
to
\begin{equation}
R_{0}=1+\left(\frac{1+(-1)^{d_{1}}}{2}d_{1}+1\right)^{2}\left(\frac{1+(-1)^{d_{2}}}{2}d_{2}+1\right)^{2}\left(\frac{1+(-1)^{d_{3}}}{2}d_{3}+1\right)^{2},\label{eq:T1D:R0:3.2}
\end{equation}
\begin{align}
R_{1}&=-(1-(-1)^{d_{1}})(d_{2}+1)(d_{3}+1),\notag\\
R_{2}&=-(1-(-1)^{d_{2}})(d_{1}+1)(d_{3}+1),\notag\\
R_{3}&=-(1-(-1)^{d_{3}})(d_{1}+1)(d_{2}+1).\label{eq:T1D:Rx:3.2}
\end{align}

Third, for two odd 
\[
d_{i}=2k+1,\quad k\in\mathbb{Z},\quad k\geqslant0
\]
and one even 
\[
d_{j}=2k,\quad k\in\mathbb{Z},\quad k\geqslant0
\]
values of the decoration multiplicity, the coefficients of the expression
for the residual entropy of the system (\ref{eq:T1D:s0}) are equal
to
\begin{equation}
R_{0}=\frac{1+(-1)^{d_{1}}}{2}(d_{1}+1)^{2}+\frac{1+(-1)^{d_{2}}}{2}(d_{2}+1)^{2}+\frac{1+(-1)^{d_{3}}}{2}(d_{3}+1)^{2},\label{eq:T1D:R0:3.3}
\end{equation}
\begin{equation}
R_{1}=R_{2}=R_{3}=0.\label{eq:T1D:Rx:3.3}
\end{equation}

In other cases, that is, for all odd values of the decoration multiplicity,
\[
d_{i}=2k+1,\quad k\in\mathbb{Z},\quad k\geqslant0,\quad i=1,\ 2,\ 3,
\]
the residual entropy is zero, and the coefficients of the expression
(\ref{eq:T1D:s0}) are respectively equal to (\ref{eq:T1D:R:0}).

4. In the absence of exchange interactions between nodal spins and
the presence of antiferromagnetic interactions between decorating
spins,
\begin{equation}
J_{i}=0,\quad J_{j}^{\prime}=-1,\quad i,j=1,\ 2,\ 3\label{eq:T1D:J0:Jm}
\end{equation}
there is only one case when the residual entropy of the system is
not equal to zero. This situation occurs when the total sum of decoration
multiplicities is even,
\[
d_{1}+d_{2}+d_{3}=2k,\quad k\in\mathbb{Z},\quad k\geqslant1,
\]
then the coefficients of the expression for the residual entropy (\ref{eq:T1D:s0})
are
\begin{equation}
R_{0}=(d_{1}+1)^{2}+(d_{2}+1)^{2}+(d_{3}+1)^{2},\label{eq:T1D:R0:4}
\end{equation}
\begin{align}
R_{1}&=(-1)^{d_{1}}2(d_{2}+1)(d_{3}+1),\notag\\
R_{2}&=(-1)^{d_{2}}2(d_{1}+1)(d_{3}+1),\notag\\
R_{3}&=(-1)^{d_{3}}2(d_{1}+1)(d_{2}+1).\label{eq:T1D:Rx:4}
\end{align}

In the case when the total sum of decoration multiplicities is odd,
\[
d_{1}+d_{2}+d_{3}=2k+1,\quad k\in\mathbb{Z},\quad k\geqslant0
\]
the residual entropy of the system is zero, and the coefficients in
the expression (\ref{eq:T1D:s0}) are respectively (\ref{eq:T1D:R:0}).

5. In the absence of exchange interactions between nodal spins and
the presence of ferromagnetic interactions between decorating spins,
\begin{equation}
J_{i}=0,\quad J_{j}^{\prime}=+1,\quad i,j=1,\ 2,\ 3,\label{eq:T1D:J0:Jp}
\end{equation}
and also in the presence of ferromagnetic interactions between nodal
and decorating spins,
\begin{equation}
J_{i}=+1,\quad J_{j}^{\prime}=+1,\quad i,j=1,\ 2,\ 3\label{eq:T1D:Jp:Jp}
\end{equation}
the residual entropy of the system is equal to zero, and the coefficients
of the expression for the residual entropy of the system (\ref{eq:T1D:s0})
are respectively equal to (\ref{eq:T1D:R:0}).

\section{Thermodynamics of a spin system with isotropic decorations}

For clarity of results, we will consider the situation when the spin
system is isotropically decorated in three directions, while the exchange
interactions are the same between all nodal spins, and also, in turn,
between all decorating spins, including decorating and nodal spins,
\begin{equation}
J_{i}=J,\quad d_{j}=d,\quad J_{j}^{\prime}=J^{\prime},\quad i,j=1,\ 2,\ 3.\label{eq:J:Jd:iso}
\end{equation}
Just such a situation is shown in Fig. \ref{fig:T1D:lattice}, which
shows a triangular lattice decorated three times in all directions.
In such an isotropic case, all the original formulas can be simplified.

The expression for the principal eigenvalue of the transfer matrix
(\ref{eq:T1D:L1}) is rewritten as
\begin{equation}
\ln\frac{\lambda}{2}=\frac{1}{8\pi^{2}(3d+1)}\intop_{0}^{2\pi}\intop_{0}^{2\pi}\ln[C^{3}+S^{3}-SD^{2}(\cos\alpha+\cos\beta+\cos(\alpha+\beta))]\:d\alpha\:d\beta,\label{eq:T1D:L1:}
\end{equation}
and the expression for spontaneous magnetization (\ref{eq:T1D:M})
becomes

\begin{equation}
M=(1-k)^{1/8},\quad k=\frac{D^{6}}{S^{3}(2C^{3}+2S^{3}+3SD^{2})},\label{eq:T1D:M:}
\end{equation}
where the corresponding coefficients 
\begin{equation}
C=\frac{1}{2}\left[e^{\frac{2J}{T}}\left(c_{d}^{1+d}+s_{d}^{1+d}\right)^{2}+e^{-\frac{2J}{T}}\left(c_{d}^{1+d}-s_{d}^{1+d}\right)^{2}\right],\label{eq:T1D:L1:C:}
\end{equation}
\begin{equation}
S=\frac{1}{2}\left[e^{\frac{2J}{T}}\left(c_{d}^{1+d}+s_{d}^{1+d}\right)^{2}-e^{-\frac{2J}{T}}\left(c_{d}^{1+d}-s_{d}^{1+d}\right)^{2}\right],\label{eq:T1D:L1:S:}
\end{equation}
\begin{equation}
D=c_{d}^{2+2d}-s_{d}^{2+2d},\label{eq:T1D:L1:D:}
\end{equation}
\[
c_{d}=\cosh\frac{J^{\prime}}{T},\quad s_{d}=\sinh\frac{J^{\prime}}{T}
\]
lose dependence on direction index~$i$.

Next, to demonstrate the most important results, we consider three
cases that clearly characterize the influence of the types of exchange
interactions between nodal and decorating spins, as well as the multiplicity
of decoration of the spin system.

1. In the first case, in the absence of competing exchange interactions,
the spin system experiences magnetic ordering, that is, the system
has a nonzero spontaneous magnetization. At zero temperature, the
system is in an equilibrium state, so the residual entropy of the
system is zero, and the residual spontaneous magnetization is saturated
and equal to unity. The system has a phase-transition temperature
($T_{\text{c}}$), above which spontaneous magnetization disappears.
The temperature dependence of the heat capacity can have a two-peak
structure. At the magnetic phase-transition point, the heat capacity
experiences a logarithmic divergence ($\lambda$-shaped Onsager peak),
and at temperatures $T<T_{\text{c}}$, a dome-shaped peak can form
under certain conditions (see discussion of this effect in Refs. \cite{Zarubin:2020,Kassan-Ogly:2019,Kassan-Ogly:2018:}).
Thus, the spin system is non-degenerate and does not experience frustration,
and the basic thermodynamic characteristics of the system have the
following values
\begin{equation}
s^{\circ}=0,\quad M^{\circ}=1,\quad T_{\text{c}}>0.\label{eq:T1D:TD:1}
\end{equation}
(Note that this behavior is also observed in the case of undecorated
lattices.)

\begin{figure}[tbh]
\centering \includegraphics[scale=1.33]{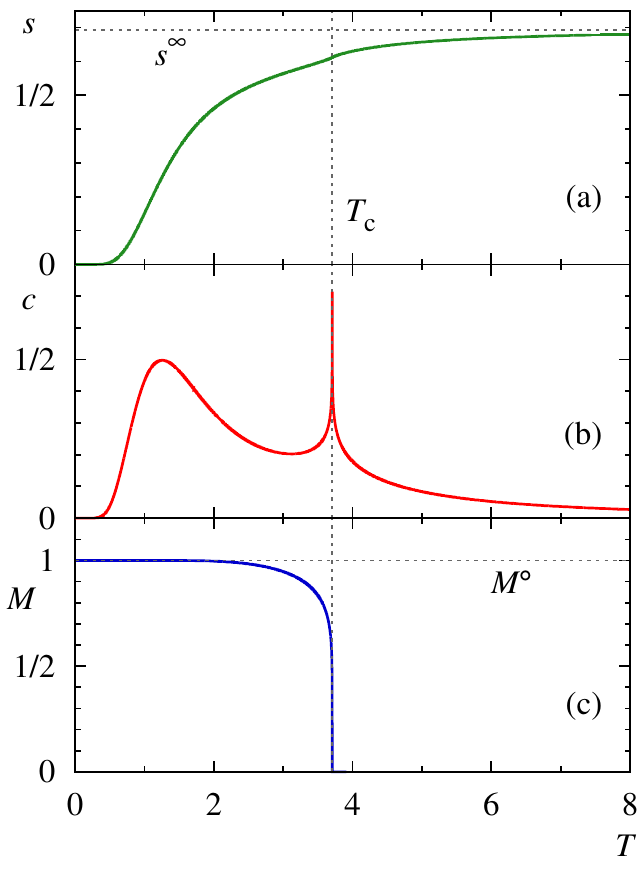}
\caption{Entropy (green line, a), heat capacity (red line, b) and spontaneous
magnetization (blue line, c) for a triangular lattice with triply
decorated ($d=3$) in all directions at ferromagnetic interaction
($J=+1$) between nodal spins and at ferromagnetic (or antiferromagnetic)
interaction ($J^{\prime}=\pm1$) between decorating spins, where $s^{\circ}=0$,
$T_{\text{c}}\approx3.705\,977$, $M^{\circ}=1$.}
\label{fig:T1D:p03p:TD}
\end{figure}

The behavior of the decorated system described here is observed in
the presence of ferromagnetic interactions between nodal spins and
between decorating spins,
\begin{equation}
J=+1,\quad J^{\prime}=+1,\label{eq:T1D:Jp:Jp:}
\end{equation}
for any values of the lattice decoration multiplicity,
\[
d=k,\quad k\in\mathbb{Z},\quad k\geqslant0.
\]
In this case, the residual entropy is zero, the residual spontaneous
magnetization is saturated and equal to unity, and the heat capacity
temperature function has a two-peak structure, where one peak is domed,
and the second $\lambda$-shaped peak experiences a logarithmic divergence
at the temperature of the magnetic phase transition ($T_{\text{ c}}$),
as shown in Fig. \ref{fig:T1D:p03p:TD}. Here the system behaves equivalently
to the system shown in Fig. \ref{fig:T1D:mmp:000:TD}.

\begin{figure}[tbh]
\centering \includegraphics[scale=1.33]{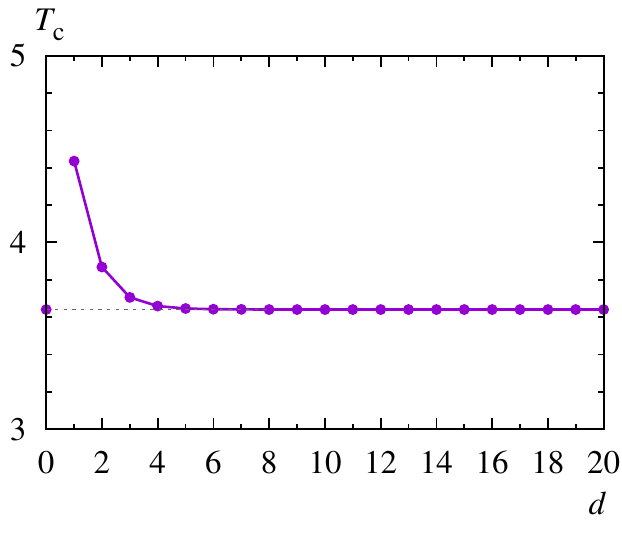}
\caption{Dependence of the magnetic phase-transition temperature ($T_{\text{c}}$)
on the decoration multiplicity ($d$) of a triangular lattice at ferromagnetic
interaction ($J=+1$) between nodal spins and at ferromagnetic interaction
($J^{\prime}=+1$) between decorating spins, where $T_{\text{c}}(d=0)=T_{\text{c}}(d=\infty)\approx3.640\,957$.}
\label{fig:T1D:pNp:Tc}
\end{figure}

For some values of the decoration multiplicity, the phase-transition
temperature has the following values 
\[
T_{\text{c}}(d=1)=\frac{2}{\ln\left(2\sqrt{3}-1\right)}\approx4.435\,439,
\]
\[
T_{\text{c}}(d=2)=\frac{4}{\ln\left(\sqrt{\frac{18-7\sqrt{3}}{2}}+\frac{3\sqrt{3}}{2}-\frac{3}{2}\right)}\approx3.868\,750,
\]
\[
T_{\text{c}}(d=3)=\frac{4}{\ln\left(2\sqrt{5-2\sqrt{3}}+2\sqrt{3}-3\right)}\approx3.705\,977,
\]
\begin{equation}
T_{\text{c}}(d=4)\approx3.658\,931,\quad T_{\text{c}}(d=5)\approx3.645\,832,\label{eq:T1D:pNp:Tc}
\end{equation}
which are shown in Fig. \ref{fig:T1D:pNp:Tc}. At the same time, it
should be noted that
\[
T_{\text{c}}(d=0)=T_{\text{c}}(d=\infty)=\frac{4}{\ln3}\approx3.640\,957.
\]

Such a situation (\ref{eq:T1D:TD:1}) can also arise in the presence
of ferromagnetic interactions between nodal spins and in the presence
of antiferromagnetic interactions between decorating spins,
\begin{equation}
J=+1,\quad J^{\prime}=-1,\label{eq:T1D:Jp:Jm:}
\end{equation}
for odd values of lattice decoration multiplicity,
\begin{equation}
d=2k+1,\quad k\in\mathbb{Z},\quad k\geqslant0,\label{eq:T1D:pNp:d1}
\end{equation}
(see Fig. \ref{fig:T1D:p03p:TD}).

In this case, the phase-transition temperature has the following values
\[
T_{\text{c}}(d=1)\approx4.435\,439,\quad T_{\text{c}}(d=3)\approx3.705\,977,
\]
\[
T_{\text{c}}(d=5)\approx3.645\,832,\quad T_{\text{c}}(d=\infty)\approx3.640\,957,
\]
which is the same as (\ref{eq:T1D:pNp:Tc}) for odd decoration multiplicity
values (\ref{eq:T1D:pNp:d1}). These data can be compared with the
graph in Fig. \ref{fig:T1D:pNp:Tc}.

\begin{figure}[tbh]
\centering \includegraphics[scale=1.33]{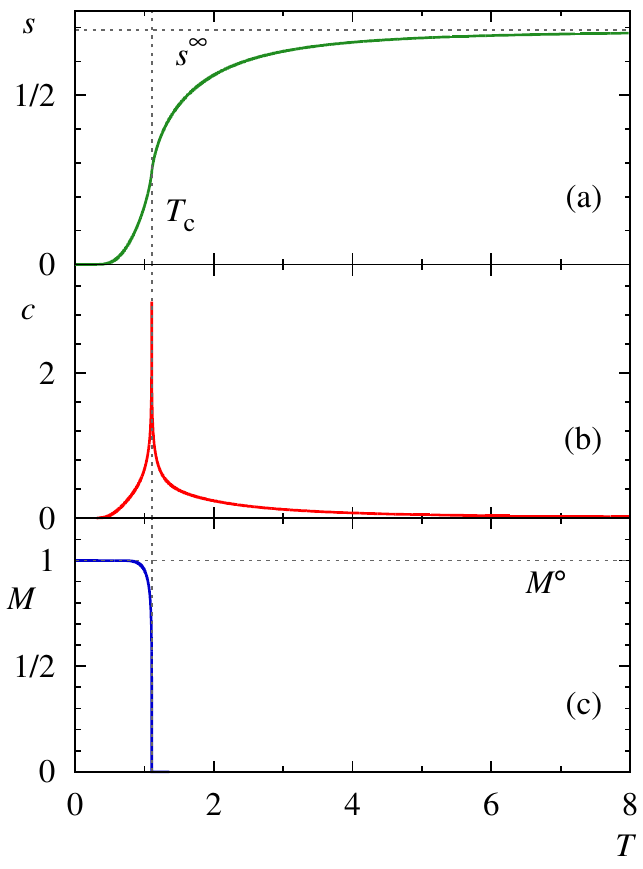}
\caption{Entropy (green line, a), heat capacity (red line, b) and spontaneous
magnetization (blue line, c) for a triangular lattice with triply
decorated ($d=3$) in all directions without exchange interactions
($J=0$) of nodal spins and at ferromagnetic (or antiferromagnetic)
interaction ($J^{\prime}=\pm1$) between decorating spins, where $s^{\circ}=0$,
$T_{\text{c}}\approx1.103\,087$, $M^{\circ}=1$.}
\label{fig:T1D:z03p:TD}
\end{figure}

It should be noted that in the absence of interactions between nodal
spins and the presence of ferromagnetic interactions between decorating
spins,
\begin{equation}
J=0,\quad J^{\prime}=+1,\label{eq:T1D:J0:Jp:}
\end{equation}
for any values of the lattice decoration multiplicity,
\[
d=k,\quad k\in\mathbb{Z},\quad k\geqslant0,
\]
the behavior of the system corresponds to the condition (\ref{eq:T1D:TD:1})
and is shown in Fig. \ref{fig:T1D:z03p:TD}.

\begin{figure}[tbh]
\centering \includegraphics[scale=1.33]{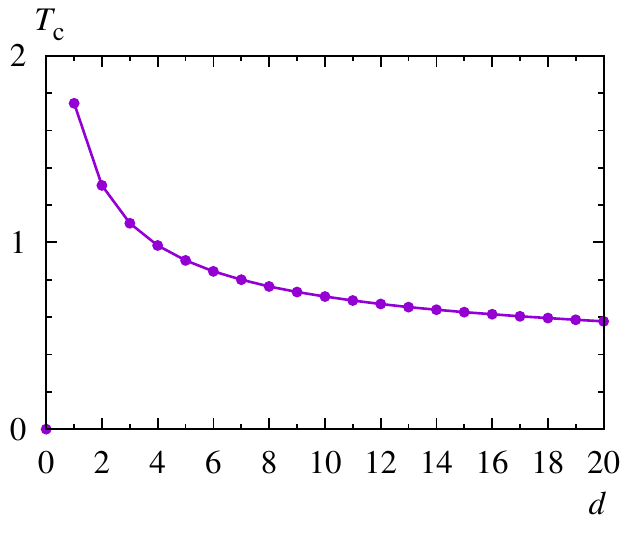}
\caption{Dependence of the magnetic phase-transition temperature ($T_{\text{c}}$)
on the decoration multiplicity ($d$) of a triangular lattice without
exchange interactions of nodal spins ($J=0$) and at ferromagnetic
interaction ($J^{\prime}=+1$) of decorating spins.}
\label{fig:T1D:zNp:Tc}
\end{figure}

In this case, the phase-transition temperatures for some values of
the decoration multiplicity, respectively, are equal to
\[
T_{\text{c}}(d=1)=\frac{2}{\ln\left(\sqrt{3}+\sqrt{2}\right)}\approx1.744\,872,
\]
\[
T_{\text{c}}(d=2)=\frac{4}{\ln\left(2\sqrt[3]{47+2\sqrt{3}}+2\sqrt[3]{47-2\sqrt{3}}+7\right)}\approx1.305\,214,
\]
\[
T_{\text{c}}(d=3)=\frac{4}{\ln\left(2\sqrt{44+18\sqrt{6}}+4\sqrt{6}+9\right)}\approx1.103\,087,
\]
\[
T_{\text{c}}(d=4)\approx0.983\,759,\quad T_{\text{c}}(d=5)\approx0.903\,511,
\]
which is seen in Fig. \ref{fig:T1D:zNp:Tc}. Note that the phase transition
is absent
\[
T_{\text{c}}(d=0)=T_{\text{c}}(d=\infty)=0,
\]
at finite temperatures in the undecorated case, as well as in the
case of infinite decoration of the spin system.

A similar situation arises in the absence of interactions between
nodal spins and the presence of antiferromagnetic interactions between
decorating spins,
\begin{equation}
J=0,\quad J^{\prime}=-1,\label{eq:T1D:J0:Jm:}
\end{equation}
for odd values of the lattice decoration multiplicity,
\[
d=2k+1,\quad k\in\mathbb{Z},\quad k\geqslant0,
\]
conditions are met for the system (\ref{eq:T1D:TD:1}), see Fig. \ref{fig:T1D:z03p:TD}.

In this case, the phase-transition temperature of the system is correspondingly
equal to
\[
T_{\text{c}}(d=1)\approx1.744\,872,\quad T_{\text{c}}(d=3)\approx1.103\,087,\quad T_{\text{c}}(d=5)\approx0.903\,511,\quad T_{\text{c}}(d=\infty)=0,
\]
which can be correlated with the results presented in Fig. \ref{fig:T1D:zNp:Tc}
for odd decoration multiplicity values.

2. In the second case, a competition of exchange interactions between
spins arises in the system, which destroys the long-range magnetic
order, while the system experiences degeneracy, which leads to a rearrangement
of the magnetic structure of the ground state, which begins to include
a set of spin configurations comparable with the size of the system,
including the absence of translational invariance. Even at zero temperature,
the degeneracy of the spin system does not disappear. Such a spin
system is in the regime of magnetic frustrations. It is the frustrations
that destroy the long-range magnetic order, while no magnetic ordering
occurs in the system even at zero temperature, that is, there is no
spontaneous magnetization, and the residual entropy is nonzero. The
system does not experience a temperature magnetic phase transition
at $T>0$, and the temperature dependence of the heat capacity has
single-domed peak that does not diverge. In this case, the basic thermodynamic
parameters of this state of the system can be written in the form
\begin{equation}
0<s^{\circ}<\ln2,\quad M=0,\quad T_{\text{c}}=0.\label{eq:T1D:TD:2}
\end{equation}
(Note that this behavior is also characteristic of undecorated lattices.)

This behavior of the system is observed in the presence of antiferromagnetic
interactions between nodal and between decorating spins,
\begin{equation}
J=-1,\quad J^{\prime}=-1,\label{eq:T1D:Jm:Jm:}
\end{equation}
a competition of exchange interactions arises in the spin system,
and the exchange interactions between decorating spins only exacerbate
this competition.

\begin{figure}[tbh]
\centering \includegraphics[scale=1.33]{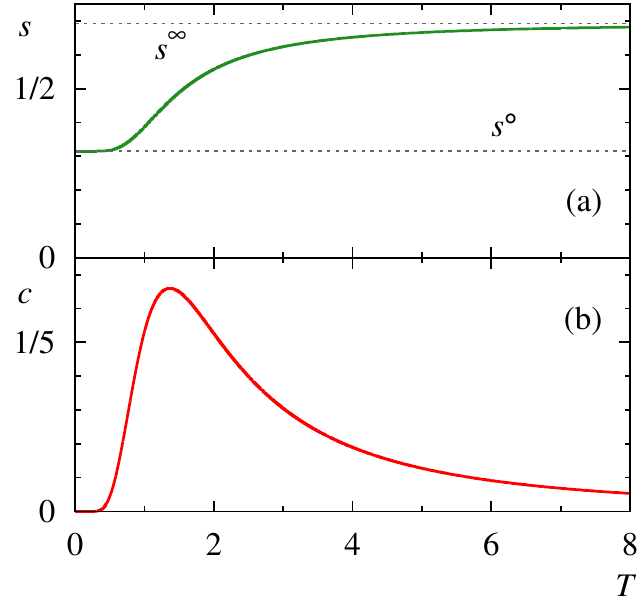}
\caption{Entropy (green line, a) and heat capacity (red line, b) for a triangular
lattice with triply decorated ($d=3$) in all directions at antiferromagnetic
interaction ($J=-1$) between nodal spins and at antiferromagnetic
(or ferromagnetic) interaction ($J^{\prime}=\mp1$) between decorating
spins, where $s^{\circ}\approx0.315\,139$, $T_{\text{c}}=0$, $M=0$.}
\label{fig:T1D:m:03m:TD}
\end{figure}

An example of the behavior of the thermodynamic functions of the spin
system in the case under consideration is shown in Fig.~\ref{fig:T1D:m:03m:TD}.
Here the system behaves equivalently to the behavior of the system
shown in Fig. \ref{fig:T1D:ppm:000:TD}.

It is important to note that in addition to the absence of spontaneous
magnetization in the system, an important characteristic is the presence
of a nonzero residual entropy. Next, we give expressions for the residual
entropy in the situation under consideration.

Thus, if the lattice decoration multiplicity is even,
\[
d=2k,\quad k\in\mathbb{Z},\quad k\geqslant0,
\]
then the residual entropy of the system is greater than zero for the
exchange interaction parameters (\ref{eq:T1D:Jm:Jm:}), and the coefficients
of the expression (\ref{eq:T1D:s0}) are rewritten to the form
\[
R_{0}=3(d+1)^{2},\quad R_{1}=R_{2}=R_{3}=2(d+1)^{2}.
\]
In this case, the expression for the residual entropy of the system
(\ref{eq:T1D:s0}) can also be expressed in terms of a single integral
as
\begin{equation}
s^{\circ}=\frac{1}{4\pi(3d+1)}\intop_{0}^{2\pi}\ln\frac{\vartheta+\sqrt{\vartheta^{2}-16(d+1)^{4}\cos^{2}\alpha}}{2}\,d\alpha,\label{eq:T1D:s0:1.1}
\end{equation}
\[
\vartheta=(d+1)^{2}+4(d+1)^{2}\cos^{2}\alpha.
\]

\begin{figure}[tbh]
\centering \includegraphics[scale=1.33]{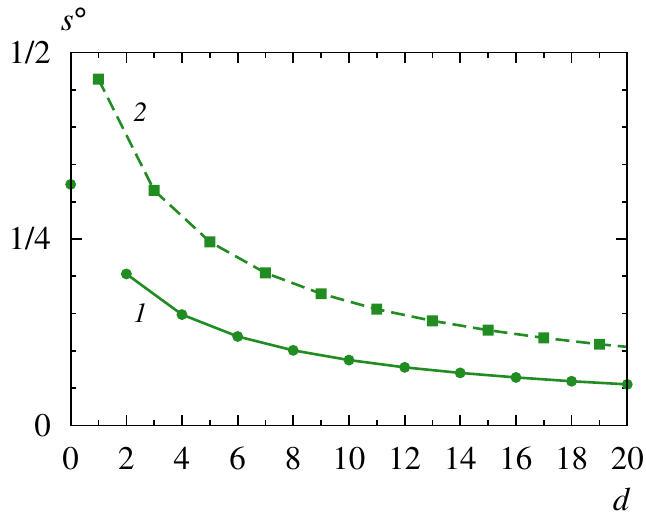}
\caption{Dependence of the residual entropy ($s^{\circ}$) on the decoration
multiplicity ($d$) of a triangular lattice at antiferromagnetic interaction
($J=-1$) between nodal spins and at antiferromagnetic interaction
($J^{\prime}=-1$) between decorating spins. Line 1 corresponds to
the values of the residual entropy at even decoration multiplicity,
and line 2 corresponds to the entropy values at odd decoration multiplicity
of a triangular lattice.}
\label{fig:T1D:m:Nm:S0}
\end{figure}

Let us write down some values of the residual entropy for some values
of the decoration multiplicity,
\[
s^{\circ}(d=0)\approx0.323\,066,\quad s^{\circ}(d=2)\approx0.203\,097,\quad s^{\circ}(d=4)\approx0.148\,654,\quad s^{\circ}(d=\infty)=0,
\]
which are shown in Fig. \ref{fig:T1D:m:Nm:S0} (line~1).

On the other hand, if the decoration multiplicity is odd,
\[
d=2k+1,\quad k\in\mathbb{Z},\quad k\geqslant0,
\]
then the coefficients (\ref{eq:T1D:R0:1.2}) and (\ref{eq:T1D:Rx:1.2})
of the expression (\ref{eq:T1D:s0}) take the form
\[
R_{0}=3(d+1)^{4}+1,\quad R_{1}=R_{2}=R_{3}=2d(d+1)^{2}(d+2).
\]

In this case, it is also possible to obtain an expression for the
residual entropy (\ref{eq:T1D:s0}) in the form of a single integral
in the form
\begin{equation}
s^{\circ}=\frac{1}{4\pi(3d+1)}\intop_{0}^{2\pi}\ln\frac{\vartheta+\sqrt{\vartheta^{2}-16d^{2}(d+1)^{4}(d+2)^{2}\cos^{2}\alpha}}{2}\,d\alpha,\label{eq:T1D:s0:1.2}
\end{equation}
\[
\vartheta=(d^{2}+2d+2)^{2}+4d(d+1)^{2}(d+2)\cos^{2}\alpha.
\]

The values of the residual entropy of the system for a number of decoration
values are given here
\[
s^{\circ}(d=1)\approx0.464\,147,\quad s^{\circ}(d=3)\approx0.315\,139,\quad s^{\circ}(d=5)\approx0.246\,059,\quad s^{\circ}(d=\infty)=0,
\]
and shown in Fig.~\ref{fig:T1D:m:Nm:S0} (line~2).

A similar situation arises in the presence of antiferromagnetic interactions
between nodal spins and in the presence of ferromagnetic interactions
between decorating spins,
\begin{equation}
J=-1,\quad J^{\prime}=+1,\label{eq:T1D:Jm:Jp:}
\end{equation}
at any value of the lattice decoration multiplicity,
\[
d=k,\quad k\in\mathbb{Z},\quad k\geqslant0,
\]
the residual entropy of the system is also greater than zero and is
equal to (\ref{eq:T1D:s0:1.2}).

In this case, for some values of lattice decoration values, the residual
entropy is equal to
\[
s^{\circ}(d=0)\approx0.323\,066,\quad s^{\circ}(d=1)\approx0.464\,147,\quad s^{\circ}(d=2)\approx0.372\,079,
\]
\[
s^{\circ}(d=3)\approx0.315\,139,\quad s^{\circ}(d=4)\approx0.275\,537,\quad s^{\circ}(d=\infty)=0.
\]

Note that the temperature behavior of the thermodynamic functions
in the cases (\ref{eq:T1D:Jm:Jm:}) and (\ref{eq:T1D:Jm:Jp:}) coincides
for odd values of the decoration multiplicity (for example, see Fig.
\ref{fig:T1D:m:03m:TD} and Fig. \ref{fig:T1D:m:Nm:S0} line~2).

Also, a similar situation can manifest itself in the absence of an
exchange interaction between nodal spins and the presence of antiferromagnetic
interactions between decorating spins (\ref{eq:T1D:J0:Jm:}), with
an even lattice decoration multiplicity,
\[
d=2k,\quad k\in\mathbb{Z},\quad k\geqslant0,
\]
here the residual entropy is also greater than zero and is equal to
(\ref{eq:T1D:s0:1.1}).

In this case, the residual entropy for some even values of the decoration
multiplicity has the values
\[
s^{\circ}(d=0)\approx0.693\,147,\quad s^{\circ}(d=2)\approx0.203\,097,\quad s^{\circ}(d=4)\approx0.148\,654,\quad s^{\circ}(d=\infty)=0.
\]

Note that the residual entropy function, depending on the value of
the decoration multiplicity, in the cases (\ref{eq:T1D:Jm:Jm:}) and
(\ref{eq:T1D:J0:Jm:}) coincides for even ($d>0$) values of the quantity
decoration multiplicity (see Fig. \ref{fig:T1D:m:Nm:S0} line~1).

It is also important to note that in all the above cases, the value
of the residual entropy tends to zero as the decoration multiplicity
increases,
\[
\lim_{d\to\infty}s^{\circ}\to0.
\]

3. In the third case, in the presence of competing exchange interactions,
magnetic frustrations and partial magnetic ordering coexist in the
spin system. It is generally accepted that in spin systems with competing
exchange interactions, frustrations can occur that suppress long-range
magnetic order; as a result, magnetic ordering does not occur in the
system and no magnetic phase transition is observed with increasing
temperature. In the situation under consideration, such coexistence
is possible, since, on the one hand, the system is frustrated and
its residual entropy is nonzero, and on the other hand, partial ordering
occurs in the system, which leads to the appearance of spontaneous
magnetization. Moreover, the spontaneous magnetization does not saturate
when approaching zero temperature, while the residual spontaneous
magnetization is less than unity. The system also experiences a magnetic
phase transition, i.e. the spontaneous magnetization is zero above
the phase-transition temperature. At the point of the magnetic phase
transition, the heat capacity function experiences a logarithmic divergence
($\lambda$-shaped Onsager peak), which corresponds to the existence
of a long-range magnetic order.

Thus, the basic thermodynamic characteristics of the system have the
following values
\begin{equation}
0<s^{\circ}<\ln2,\quad0<M^{\circ}<1,\quad T_{\text{c}}>0.\label{eq:T1D:TD:3}
\end{equation}
(Note that this is a completely new phenomenon and this behavior is
not typical of undecorated lattices.)

\begin{figure}[tbh]
\centering \includegraphics[scale=1.33]{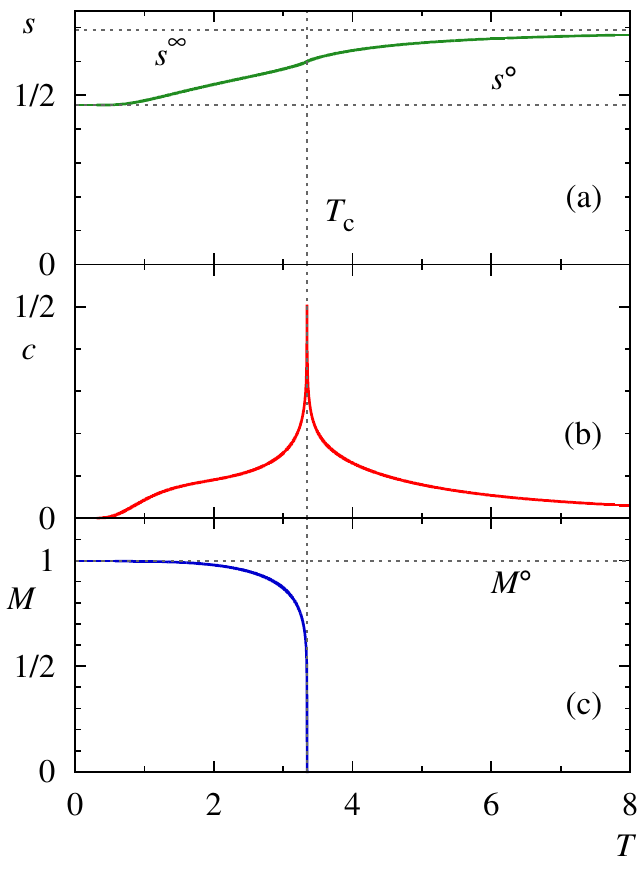}
\caption{Entropy (green line, a), heat capacity (red line, b) and spontaneous
magnetization (blue line, c) for a triangular lattice doubly decorated
($d=2$) in all directions at ferromagnetic interaction ($J=+1$)
between nodal spins and at antiferromagnetic interaction ($J^{\prime}=-1$)
between decorating spins, where $s^{\circ}\approx0.471\,630$, $T_{\text{c}}\approx3.342\,338$,
$M^{\circ}\approx0.997\,040$.}
\label{fig:T1D:p:02m:TD}
\end{figure}

In the considered isotropic variant (\ref{eq:J:Jd:iso}), in the presence
of ferromagnetic interactions between nodal spins and in the presence
of antiferromagnetic interactions between decorating spins (\ref{eq:T1D:Jp:Jm:}),
but with even values of the lattice decoration multiplicity,
\[
d=2k,\quad k\in\mathbb{Z},\quad k\geqslant0,
\]
the residual entropy is greater than zero. The behavior of thermodynamic
functions is shown in Fig. \ref{fig:T1D:p:02m:TD}.

In the case under discussion, the coefficients of the expression for
the residual entropy (\ref{eq:T1D:s0}) are
\[
R_{0}=(d+1)^{2}(d^{4}+4d^{3}+6d^{2}+4d+4),\quad R_{1}=R_{2}=R_{3}=-2d(d+1)^{2}(d+2).
\]
Moreover, this expression can also be expressed in terms of a single
integral as
\[
s^{\circ}=\frac{1}{4\pi(3d+1)}\intop_{0}^{2\pi}\ln\frac{\vartheta+\sqrt{\vartheta^{2}-16d^{2}(d+1)^{4}(d+2)^{2}\cos^{2}\alpha}}{2}\,d\alpha,
\]
\[
\vartheta=(d+1)^{2}(d^{4}+4d^{3}+4d^{2}+4)+4d(d+1)^{2}(d+2)\cos^{2}\alpha.
\]

\begin{figure}[tbh]
\centering \includegraphics[scale=1.33]{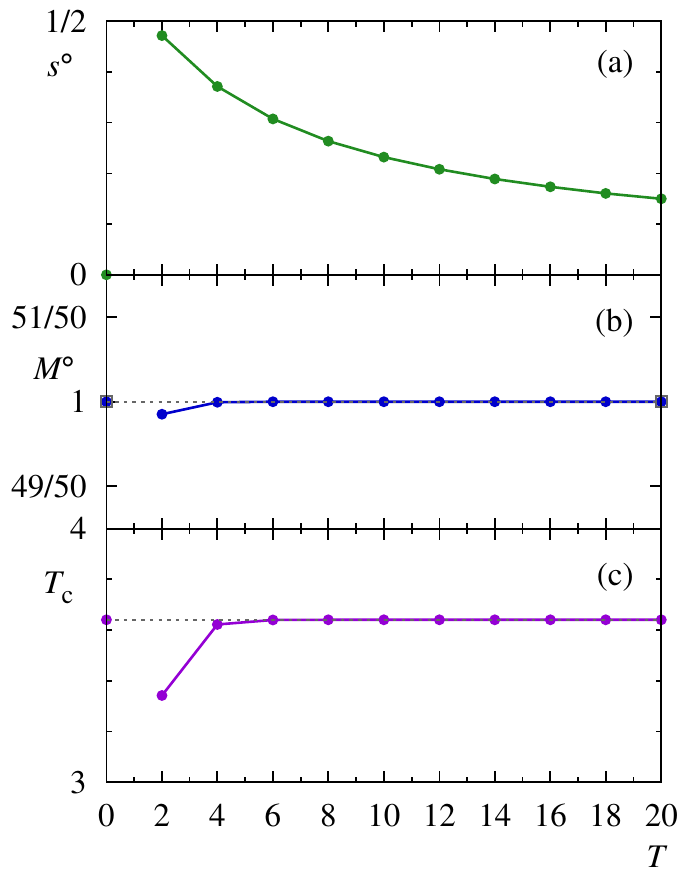}
\caption{Dependence of the residual entropy (green line, a), residual magnetization
(blue line, b) and the magnetic phase-transition temperature (magenta
line, c) on the triangular lattice decoration multiplicity at ferromagnetic
interaction ($J=+1$) between nodal spins and at antiferromagnetic
interaction ($J^{\prime}=-1$) between decorating spins, where $M^{\circ}(d=0)=M^{\circ}(d=\infty)=1$,
$T_{\text{c}}(d=0)=T_{\text{c}}(d=\infty)\approx3.640\,957$.}
\label{fig:T1D:p:Nm:TD}
\end{figure}

The values of the residual entropy of the system for a number of even
values of decoration values are given below
\[
s^{\circ}(d=2)\approx0.471\,630,\quad s^{\circ}(d=4)\approx0.371\,431,\quad s^{\circ}(d=6)\approx0.307\,251,\quad s^{\circ}(d=\infty)=0,
\]
and are shown in Fig. \ref{fig:T1D:p:Nm:TD}a. Note that there is
no residual entropy at the limiting values of the decoration multiplicity,
\[
s^{\circ}(d=0)=s^{\circ}(d=\infty)=0.
\]

It is important to note here that the spontaneous magnetization at
zero temperature does not experience saturation, that is, $M^{\circ}<1$.
And in the case under consideration, we have the values of the residual
spontaneous magnetization depending on the values of the lattice decoration
multiplicity
\[
M^{\circ}=(1-k^{\circ})^{1/8},
\]
\begin{multline*}
k^{\circ}=32(d_{1}+1)^{2}(d_{2}+1)^{2}(d_{3}+1)^{2}\\
\times\{2[(d_{1}+1)^{2}d_{2}^{2}(d_{2}+2)^{2}d_{3}^{2}(d_{3}+2)^{2}+d_{1}^{2}(d_{1}+2)^{2}(d_{2}+1)^{2}d_{3}^{2}(d_{3}+2)^{2}\\
+d_{1}^{2}(d_{1}+2)^{2}d_{2}^{2}(d_{2}+2)^{2}(d_{3}+1)^{2}]+d_{1}^{2}(d_{1}+2)^{2}d_{2}^{2}(d_{2}+2)^{2}d_{3}^{2}(d_{3}+2)^{2}\\
+d_{1}(d_{1}+2)(d_{1}^{2}+2d_{1}+2)d_{2}(d_{2}+2)(d_{2}^{2}+2d_{2}+2)d_{3}(d_{3}+2)(d_{3}^{2}+2d_{3}+2)\sign J_{1}\sign J_{2}\sign J_{3}\}^{-1}.
\end{multline*}
And in the isotropic case of residual spontaneous magnetization, depending
on the value of the multiplicity of the isotropic decoration of the
lattice, it is already equal to
\begin{equation}
M^{\circ}=(1-k^{\circ})^{1/8},\quad k^{\circ}=\frac{16(d+1)^{2}}{d^{3}(d+2)^{3}\left(d^{2}+2d+4\right)}.\label{eq:T1D:M0:iso}
\end{equation}
In this case, the values of spontaneous magnetization (\ref{eq:T1D:M0:iso})
for some values of the decoration multiplicity are equal to 
\[
M^{\circ}(d=2)\approx0.997\,040,\quad M^{\circ}(d=4)\approx0.999\,871,
\]
\[
M^{\circ}(d=6)\approx0.999\,983,\quad M^{\circ}(d=8)\approx0.999\,996,
\]
and are shown in Fig. \ref{fig:T1D:p:Nm:TD}b. It can be seen that
the residual spontaneous magnetization is maximum and equal to unity
only for the following values of the decoration multiplicity
\[
M^{\circ}(d=0)=M^{\circ}(d=\infty)=1,
\]
that is, in the case without decoration and in the case of an infinite
decoration with the spins of the triangular lattice in all directions.

The spontaneous magnetization vanishes above the phase-transition
temperature ($T_{\text{c}}$), and the heat capacity at this point
experiences a logarithmic divergence, as shown in Fig. \ref{fig:T1D:p:02m:TD}.

The values of the phase-transition temperatures of the magnetic phase
transition at some values of the decoration multiplicity are equal
to 
\[
T_{\text{c}}(d=2)=\frac{4}{\ln\left(\frac{4}{\sqrt{3}}+1\right)}\approx3.342\,338,\quad T_{\text{c}}(d=4)\approx3.622\,290,\quad T_{\text{c}}(d=6)\approx3.639\,640,
\]
and shown in Fig. \ref{fig:T1D:p:Nm:TD}c. Also note that
\[
T_{\text{c}}(d=0)=T_{\text{c}}(d=\infty)=\frac{4}{\ln3}\approx3.640\,957,
\]
The phase-transition temperature is finite and greater than zero in
the case without decoration and in the case of an infinite decoration
in all directions of the triangular lattice.

Obviously, in the last third case, frustrations are present, but they
are not capable of suppressing the magnetic ordering of the spin system,
as in the first case. This behavior is associated with the presence
of partial magnetic ordering in the system, which was considered in
detail in Ref.~\cite{Kassan-Ogly:2022}.

\section{Conclusions}

The paper studies the frustration properties of the Ising model on
a decorated triangular lattice.

For the first time, analytical expressions are obtained for the entropy,
heat capacity, and spontaneous magnetization of such a system, as
well as for the residual (zero-temperature) entropy and residual (zero-temperature)
spontaneous magnetization in the cases of anisotropic and isotropic
decoration of the spin system on a triangular lattice.

The influence of the types of exchange interactions of spins in the
nodal and decorated positions of the lattice and their ratios, as
well as the decoration multiplicity on the formation of magnetic frustrations
and the behavior of the thermodynamic and magnetic characteristics
of a decorated triangular lattice, are studied.

It is shown that it is the presence of competing exchange interactions
between spins that leads to the formation of frustrated states of
the system and also affects the magnetic ordering of the spin system.

It is also shown that despite significant simplifications in the isotropic
case, when the lattice decoration multiplicities are the same in all
directions, and the values of exchange interactions (both between
nodal and decorating spins) are equal in absolute value to unity,
spin systems can experience several regimes, during the formation
of which magnetic frustrations play an important role.

In the first case, situations are shown where, at certain ratios of
the model parameters, there are no competing exchange interactions,
which leads to the absence of magnetic frustrations, while the residual
entropy is zero, and the spin system experiences a magnetic phase
transition at the point at which the heat capacity has a logarithmically
divergent $\lambda$-shaped Onsager peak. In turn, the residual spontaneous
magnetization is saturated and equals unity, and at the point of the
magnetic phase transition, the spontaneous magnetization vanishes.

In the second case, the frustrations arising due to competing exchange
interactions completely suppress the magnetic long-range order, and
no magnetic phase transition occurs in the system with a change in
temperature. In this case, there is no spontaneous magnetization,
the temperature dependence of the heat capacity has single-domed peak,
and the residual entropy is nonzero.

In the third situation, the decorated spin system (in the presence
of competing exchange interactions) also experiences frustrations,
while the residual entropy is not equal to zero, but the spin system
has partial magnetic ordering, which leads to a phase transition at
the point where the temperature dependence of the heat capacity has
a $\lambda$-shaped Onsager peak.

It is important to note that this is a completely new phenomenon discovered
by us, namely, the coexistence of long-range magnetic order and magnetic
frustrations in a decorated lattice.

Thus, depending on the absence of competing exchange interactions
or their presence (which lead to complete magnetic disorder or to
partial magnetic ordering of the spin system), the model allows one
to describe different types of behavior of entropy, heat capacity,
and spontaneous magnetization.

The results of the study are illustrated by the observed dependences
of the residual entropy, residual magnetization, and of the magnetic
phase-transition temperature on the decoration multiplicity of the
triangular lattice in the isotropic case.

We emphasize that, contrary to the prevalent belief that frustration
(non-zero residual entropy) and phase transition (logarithmic divergence
of heat capacity) are necessarily mutually exclusive features, we
found that although this is often and often true, the frustrations
and phase transition can coexist. We note that such coexistence is
possible in the discovered regime, since the frustrations of the spin
system are not capable of completely suppressing the long-range magnetic
order, which leads to a partial magnetic ordering of the spin system.
As a result, the spontaneous magnetization does not reach saturation
at zero temperature and is less than unity, while the spontaneous
magnetization disappears at the point of the magnetic phase transition.

In conclusion, we note that the cases considered in the paper do not
fully describe all the situations that can be obtained within the
framework of the presented model. The model used and the exact analytical
solutions obtained in this paper have a great potential for describing
and understanding forused systems on a decorated triangular lattice.

\section*{Acknowledgment}

The research was carried out within the state assignment of Ministry of Science and Higher Education of the Russian Federation (theme ``Quantum'' No. 122021000038-7).

\end{document}